\documentclass[manuscript=letter,layout=twocolumn]{achemso}
\usepackage{mathtools}
\usepackage{units}
\usepackage{amssymb}
\usepackage{amsmath}
\usepackage{amsthm}
\usepackage{graphicx} 
\usepackage{subcaption}
\usepackage[usenames,dvipsnames,svgnames,table]{xcolor}

\newcommand{\red}[1]{\textcolor{black}{#1}}
\newcommand{\blue}[1]{\textcolor{black}{#1}}
\newcommand{\nela}{\langle N_{\rm ela} \rangle}

\title{An alternative approach for the determination of mean free paths of 
       electron scattering in liquid water based on experimental data}

\author{Axel Schild}
\affiliation{ETH Z\"urich,  Laboratorium f\"ur Physikalische Chemie,  8093 Z\"urich, Switzerland}
\email{axel.schild@phys.chem.ethz.ch}

\author{Michael Peper}
\affiliation{ETH Z\"urich,  Laboratorium f\"ur Physikalische Chemie,  8093 Z\"urich, Switzerland}

\author{Conaill Perry}
\affiliation{ETH Z\"urich,  Laboratorium f\"ur Physikalische Chemie,  8093 Z\"urich, Switzerland}

\author{Dominik Rattenbacher}
\affiliation{ETH Z\"urich,  Laboratorium f\"ur Physikalische Chemie,  8093 Z\"urich, Switzerland}
\alsoaffiliation{Max Planck Institute for the Science of Light, Staudtstr. 2, 91058 Erlangen, Germany}

\author{Hans Jakob W\"orner}
\affiliation{ETH Z\"urich,  Laboratorium f\"ur Physikalische Chemie,  8093 Z\"urich, Switzerland}

\begin{document}

  \begin{abstract}
    The mean free paths of low-energy electrons in liquid water are of fundamental 
    importance for modelling radiation damage and many related physico-chemical 
    processes. 
    Neither theoretical predictions nor experimental estimations have so far 
    converged to yield reliable values for these parameters. 
    We therefore introduce a new approach to determine the elastic and 
    inelastic mean free paths (EMFP, IMFP) \red{based on experimental data}. 
    We report extensive ab-initio 
    calculations of electron quantum scattering with water clusters, which are 
    brought to convergence with respect to the cluster size. This provides a 
    first-principles approach to condensed-phase scattering that includes both 
    multiple-scattering and condensation effects. The obtained differential 
    cross sections are used in a detailed Monte-Carlo simulation to extract 
    EMFP and IMFP from two recent liquid-microjet experiments that determined 
    the effective attenuation length (EAL) and the photoelectron angular 
    distribution (PAD) following oxygen 1s-ionization of liquid water.
    For electron kinetic energies from \unit[10]{eV} to \unit[300]{eV}, we 
    find that the IMFP is noticeably larger than the EAL.
    \red{The EMFP is longer than that of gas-phase water and the IMFP is 
    longer compared to the latest theoretical estimations, but both the EMFP and 
    IMFP are much shorter than suggested by experimental results for amorphous ice.} 
    The Python module developed for the analysis is available at 
    \url{https://gitlab.com/axelschild/CLstunfti} and can be used to further 
    refine our results when new experimental data become available.
  \end{abstract}
  
  \maketitle
  
  \section{Article}
  
  Knowledge of the scattering properties of electrons in liquid water is 
  vital for understanding and for modeling many physical processes, for 
  example the effect of radiation damage in living tissue 
  \cite{wang2009,kumar2010,nikjoo2016}.
  To accurately model the interaction of electrons with water molecules,
  e.g.\ with Monte-Carlo simulations \cite{champion2003,shin2018}, the differential 
  scattering cross sections (DCS) as well as the total cross sections for many 
  possible types of interactions are needed.
  Those quantities are experimentally accessible for gas-phase water \cite{katase1986,cho2004,khakoo2008,khakoo2013}, 
  but for water in the liquid phase they are difficult to obtain.
  
  Instead of the cross sections, the elastic mean free path
  (EMFP) and the inelastic mean free path (IMFP) for a given electron kinetic 
  energy (eKE) can be used as a measure for the probability to scatter.
  The IMFP is a particularly relevant parameter because it represents the 
  effective path length that an electron can travel in liquid water before it 
  scatters inelastically, thereby losing some of its energy and thus changing 
  the probability of damaging solvated molecules.
  Hence, there has been much interest in determining the IMFP theoretically as 
  well as experimentally:
  On the one hand, there have been a number of theoretical estimations of the IMFP 
  \cite{ashley1988,tomita1997,dingfelder1998,akkerman1999,pimblott2002,
  emfietzoglou2005,emfietzoglou2007,emfietzoglou2012,emfietzoglou2015,
  shinotsuka2017,garcia2017,emfietzoglou2017,nguyen2018}.
  Especially for eKE below \unit[100]{eV}, the often
  utilized models, which are based on the dielectric function, have limitations 
  \red{that need to be overcome.
  These include, e.g., the neglect of electron-exchange effects and an 
  overestimation of the needed energy-loss function in the presence of an energy
  gap in the excitation spectrum
  (see \citet{emfietzoglou2015} and \citet{nguyen2018} for details)}.
  On the other hand, the mean free path is a quantity that eludes direct 
  measurements.
  What has been measured directly, however, are the effective attenuation lengths
  (EAL) of electrons in liquid water \cite{ottosson2010,buchner2012,suzuki2014} 
  and the photoelectron angular distributions (PAD) of electrons generated 
  through core-level ionization inside liquid water \cite{thurmer2013}.
  While the EAL provides a lower bound for the IMFP, the measured PAD of liquid 
  water was used to estimate the IMFP directly with a simplified model.
  It was found that for an eKE below \unit[$\approx 100$]{eV}, the IMFP appears 
  to be considerably shorter and flatter than previously assumed \cite{thurmer2013}. 
  \blue{
  A recent improvement of the theoretical algorithms used to determine the IMFP 
  within the dielectric formalism yielded results that qualitatively agree with 
  the experimental results\cite{nguyen2018}.}
  
  \red{
  There have also been attempts to determine scattering cross sections 
  from photoelectron imaging of water droplets \cite{signorell2016} with the 
  surprising finding that the IMFP  and scattering cross sections\cite{luckhaus2017}
  were identical to those measured for amorphous ice \cite{michaud2003} within 
  the quoted accuracies.
  In a subsequent work,\cite{gartmann2018} attempts were made to explain 
  angle-resolved photoelectron-imaging results for water clusters with the cross 
  sections, but they had to be rescaled to qualitatively reproduce the 
  experimental results.
  Discussions in the literature\cite{shinotsuka2017,bartels2019} show that it 
  is thus appropriate to conclude that there is no consensus regarding the EMFP 
  and IMFP values for electron scattering in liquid water, neither from 
  theoretical nor from experimental approaches. 
  This situation motivates the present work.}
  
  In the following, we present an alternative approach to determine the EMFP 
  and IMFP of electron scattering in liquid water \red{based on experimental 
  data}. 
  Our approach relies on extensive \textit{ab-initio} quantum-scattering 
  calculations of electrons with water clusters of variable size to achieve a 
  first-principles description of electron scattering in liquid water. 
  The approach includes, at a quantum-mechanical level, the key characteristics 
  that distinguish electron scattering in the gas and condensed phases. 
  These differences originate from (i) multiple-scattering and interference 
  effects originating from scattering at neighboring molecules in the condensed 
  phase and (ii) the modification of the electronic structure of the water 
  molecules through condensation. Verifying the convergence of our scattering 
  calculations with respect to cluster size ensures that the relevant effects 
  have been taken into account.

  We then use Monte-Carlo electron-trajectory calculations to connect these 
  {\it ab-initio} results with experimental observables. 
  To uniquely determine EMFP and IMFP values which are consistent with 
  experimental observations, we need two independent measurements. 
  For this purpose, we chose the measurements of the EAL by \citet{suzuki2014} 
  and of the PAD by \citet{thurmer2013}, both realized by ionizing the oxygen 
  1s-orbital of liquid water in microjets. 
  We use classical-trajectory Monte-Carlo simulations to describe the electron 
  transport. 
  Such calculations are expected to be valid for kinetic energies above 
  ca.\ \unit[$10$]{eV}, 
  where quantum-interference effects have been shown to become 
  negligible\cite{liljequist2008}.
  In our approach, the only free parameters are the EMFP and the IMFP, while all 
  other input parameters are determined from \emph{ab initio} calculations or 
  from the measurements.
  The assumptions made in the Monte-Carlo simulations are described in the 
  supplementary material.
  
  The EMFP and IMFP are defined by assuming that for both elastic and inelastic 
  scattering, the probability of not having scattered until a distance $r$ is
  \begin{align}
    P(r) = \frac{1}{r_{\rm MFP}} e^{-\nicefrac{r}{r_{\rm MFP}}}.
    \label{eq:mfp}
  \end{align}
  Then, $\int\limits_0^{\infty} r P(r) dr = r_{\rm MFP}$
  is the corresponding mean free path.

  The EAL is a distance parameter that describes the exponential decay
  of the number of electrons $S(z)$ detected outside the liquid,
  \begin{align}
    S(z) \propto \exp(-z/r_{\rm EAL}),
    \label{eq:eal}
  \end{align}
  as a function of the distance $z$ from their point of creation to the surface.
  It is measured for an eKE corresponding to the kinetic energy after ionization.
  Since the depth of the conduction band (the electron affinity) of liquid water 
  is well below \unit[1]{eV},\cite{coe1997,chen2016,gaiduk2018,bartels2019} 
  we neglect it  due to the comparably high eKE \unit[$\ge 10$]{eV} that we consider.
  The EAL is related to the IMFP, because inelastic scattering is 
  responsible for the loss of signal $S(z)$ with starting depth $z$ of the 
  electron, and the EAL is a lower bound for the IMFP.
  
  For an isotropic sample ionized with linearly polarized light, the PAD 
  relative to the direction of polarization is given by \cite{reid2003}
  \begin{align}
    \text{PAD}(\theta) \propto 1 + \beta P_2(\cos \theta)
  \end{align}
  where $\theta$ is the polar angle, $P_2$ is the Legendre polynomial of second 
  order, and $\beta$ is the asymmetry parameter.
  The PAD of photoemission from a dense medium is closely related to the mean 
  number of elastic collisions before inelastic scattering occurs,
  \begin{align}
    \nela = \frac{r_{\rm IMFP}}{r_{\rm EMFP}},
    \label{eq:nela}
  \end{align}
  because each elastic collision changes the angular distribution of the 
  electrons at the observed eKE.
  
  The DCS for electronically, vibrationally, and rotationally elastic 
  ($J=0 \rightarrow J'=0$) scattering are computed with the program ePolyScat
  \cite{gianturco1994,natalense1999} for water clusters of different sizes and 
  shapes with nuclear configurations taken from \citet{temelso2011}, \red{and also 
  for some structures based on the modeling of experimental data for the 
  first solvation shell of liquid water.\cite{wernet2004}}
  ePolyScat solves the variational Schwinger equation using single-center 
  expansions, and we use molecular-orbital data from Hartree-Fock calculations 
  with a cc-pVTZ basis set obtained with the program Gaussian \cite{gaussian09}
  as input.
  
  The most important result from these calculations is that the DCS rapidly 
  converge with the number of water 
  molecules in the cluster (see Fig.\ S1 of the supplemental material).
  With increasing kinetic energy, the DCS are found to converge more rapidly as 
  a function of cluster size.
  These observations can be intuitively explained by relating the de-Broglie 
  wavelength $\lambda_{\rm dB}$ of the scattering particle to the extension of 
  the target. 
  Whereas the molecular-level description of a 10~eV electron 
  ($\lambda_{\rm dB}\approx 3.9$\AA) collision requires a cluster of 6-7 water 
  molecules to reach convergence, a 50~eV electron
  ($\lambda_{\rm dB}\approx 1.7$\AA) collision is well described by a much 
  smaller cluster.
  The observed rapid convergence of the DCS suggests that such cluster 
  calculations can be a good approximation for the DCS of bulk liquids.\cite{blanco2014} 
  In the present work, we used the DCS of the largest considered cluster, 
  (H$_2$O)$_7$, to describe the DCS of bulk water.
  
  The Monte-Carlo trajectory calculations were realized with the Python module 
  \texttt{CLstunfti}, which is available at \url{https://gitlab.com/axelschild/CLstunfti}.
  Here, we only give a brief overview of the computational model:
  Electrons at a given eKE are represented by classical trajectories and
  a set of EMFP and IMFP are chosen for the simulation.
  To simulate ionization, an ionization depth $z$ is selected 
  and initial directions for the trajectories are sampled from the experimental 
  gas-phase PAD.
  To account for inelastic scattering, each trajectory is given a random 
  maximum path length to travel according to the distribution \eqref{eq:mfp} 
  and using $r_{\rm MFP} = r_{\rm IMFP}$.
  It is assumed that inelastic scattering happens at this distance and that if 
  this happens, the electron is no longer detected within the relevant eKE 
  range.
  This separation of scattering into elastic and inelastic events is based on 
  the experimental observation of well separated groups of primary and secondary 
  electrons. 
  Only the primary electrons, which form a well-defined Gaussian distribution 
  of $\sim$1.5~eV width are detected in the simulated photoelectron experiments. 
  The secondary electrons, which have experienced one or more electronically 
  inelastic scattering events corresponding to an energy loss of 7~eV or more, 
  are not considered.
  To propagate the trajectories after the ionization event, a travel distance is 
  selected randomly for each trajectory according to the distribution 
  \eqref{eq:mfp} with $r_{\rm MFP} = r_{\rm EMFP}$.
  If the trajectory ends outside the liquid (the liquid surface is located 
  at $z=0$, i.e., it is flat) or if its path is longer 
  than the maximum path length for the trajectory (i.e., when inelastic 
  scattering would happen), it is counted to the measured signal or it is 
  discarded, respectively.
  If this is not the case, elastic scattering happens, the position of the 
  trajectory is updated and a new direction is chosen according to the DCS.
  This last step is repeated until all trajectories have either reached the 
  surface (contributing to the measured signal) or have been discarded due 
  to their path being longer than the pre-selected maximum path.
  
  To simulate the EAL measurement, the ionizing beam is assumed to have 
  a polarization vector that is directed to the surface and the number of 
  escaped electrons is counted depending on the chosen starting depth $z$ of 
  the trajectories.
  For the simulation of the PAD of the liquid phase, random starting depths are 
  chosen.
  As done in the experiment, the direction of the polarization vector of the 
  ionizing beam is varied.
  In both experiments, the escaped electrons are counted if they exit with a 
  small polar angle.
  Ideally, the detection angle should be very small.
  In the simulations we chose it to be \unit[1]{$^{\circ}$}, but the results 
  discussed in the following are not very sensitive to this angle as long as it 
  is chosen to be small enough, and test calculations with opening angles of 
  \unit[15]{$^{\circ}$} (which are approximately the experimental opening 
  angles) yield similar numerical results compared to the ones presented below.
  We also use importance sampling with an exponential distribution for the 
  initial starting depths of the trajectories to make the simulation more 
  efficient, as trajectories starting deep inside the liquid have a low 
  probability of reaching the surface due to inelastic scattering.
  
  \begin{figure*}[htbp]
    \includegraphics[width=0.99\textwidth]{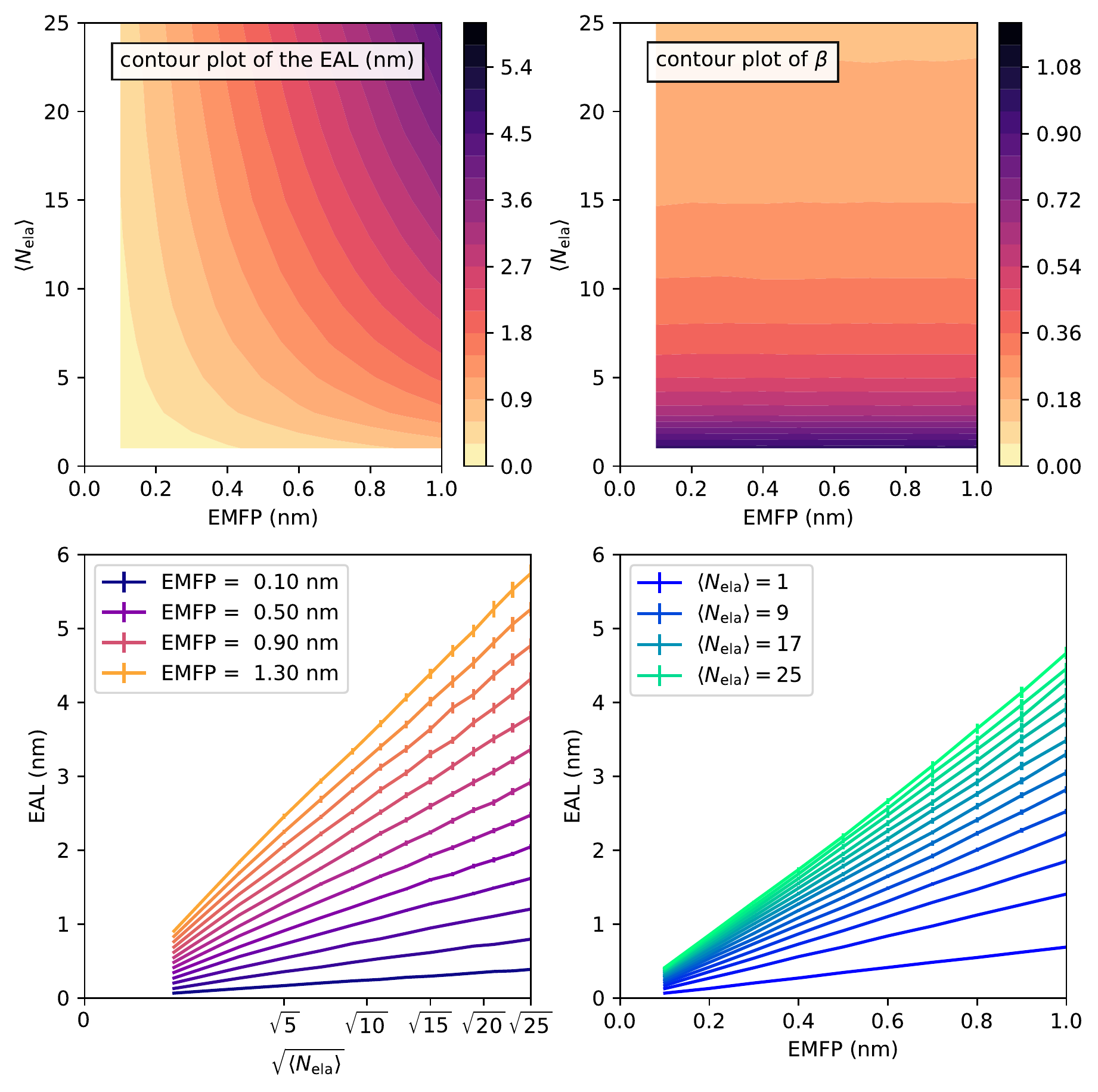}
    \caption{
    Top: Contour plots of the effective attenuation length EAL (left) and 
    the $\beta$-parameter describing the photoelectron angular distribution
    of the liquid (right) depending on the value of the elastic mean free 
    path (EMFP) and the average number of elastic collisions $\nela$.
    Bottom: Line plots of the EAL depending on $\sqrt{\nela}$ for fixed 
    EMFP (left) and depending on the EMFP for fixed $\nela$ (right).
    Small vertical lines indicate the standard deviation of the fit to 
    \eqref{eq:eal}.
    The plots show results for an electron kinetic energy of \unit[20]{eV}
    and the differential scattering cross sections obtained with ePolyScat
    for $({\rm H}_2{\rm O})_7$ clusters.
    }
    \label{fig:beh20}
  \end{figure*}
  
We start the discussion of the results with those observations that we expect to have a general character and could thus be useful in interpreting other experimental results.   To the best of our knowledge, these dependencies have not been reported before.
  Figure\ \ref{fig:beh20} illustrates the dependence of the experimental observables 
  EAL and $\beta$ on the EMFP and the average number of elastic scatterings 
  $\nela$ for an eKE of \unit[20]{eV}.
  We recall that $\nela$ encodes the dependence on the IMFP via \eqref{eq:nela}.
  \red{
  We find that 
  {\bf (i)}   for fixed EMFP  the EAL is increasing as $\sqrt{\nela}$, 
  {\bf (ii)}  for fixed EMFP  the parameter $\beta$ is decreasing as $1/\nela$ within the investigated range (top and bottom-left panels of Fig. \ref{fig:beh20}), 
  {\bf (iii)} for fixed $\nela$ the EAL is proportional to the EMFP, and 
  {\bf (iv)}  for fixed $\nela$ the parameter $\beta$ is independent of the EMFP  (top and bottom-right panels of Fig. \ref{fig:beh20}).}
  The reasons for these dependencies are the following:
  The (in)dependencies {\bf (iii)} and {\bf (iv)} on the EMFP can be explained by 
  a scaling argument:
  For fixed $\nela$, it follows from \eqref{eq:nela} that 
  trajectories from a simulation with $r_{\rm EMFP} = r_2$ are equivalent to 
  those for $r_{\rm EMFP} = r_1$ if scaled by the factor $r_2/r_1$.
  The scaling leaves the final PAD invariant and changes the depth distribution 
  from which the measured trajectories originate linearly with the EMFP.
  The linear dependence {\bf (i)} of the EAL on $\sqrt{\nela}$ is due to 
  scattering of trajectories into other directions than the forward direction -- 
  the EAL is for the trajectories like the root mean square translation 
  distance of a random walk, which also scales like $\sqrt{n}$ where $n$ is the 
  number of steps.\cite{sethna2006}
  All dependencies are monotonic, hence a set of values for the EAL and $\beta$
  corresponds to a unique set of values for the EMFP and IMFP.
These dependencies also illustrate a general protocol for analyzing the experimental results. The measurement of $\beta$ uniquely defines $\nela$, i.e. the ratio of IMFP to EMFP. Knowing $\nela$, the measurement of the EAL uniquely defines the EMFP, which in turn defines the IMFP. As a consequence, an overestimation of the EAL would result in EMFP and IMFP values that are both too long by the same factor. Errors in the measurement of $\beta$ would affect the ratio of IMFP to EMFP.
  
  \begin{figure*}[htbp]
    \includegraphics[width=0.7\textwidth]{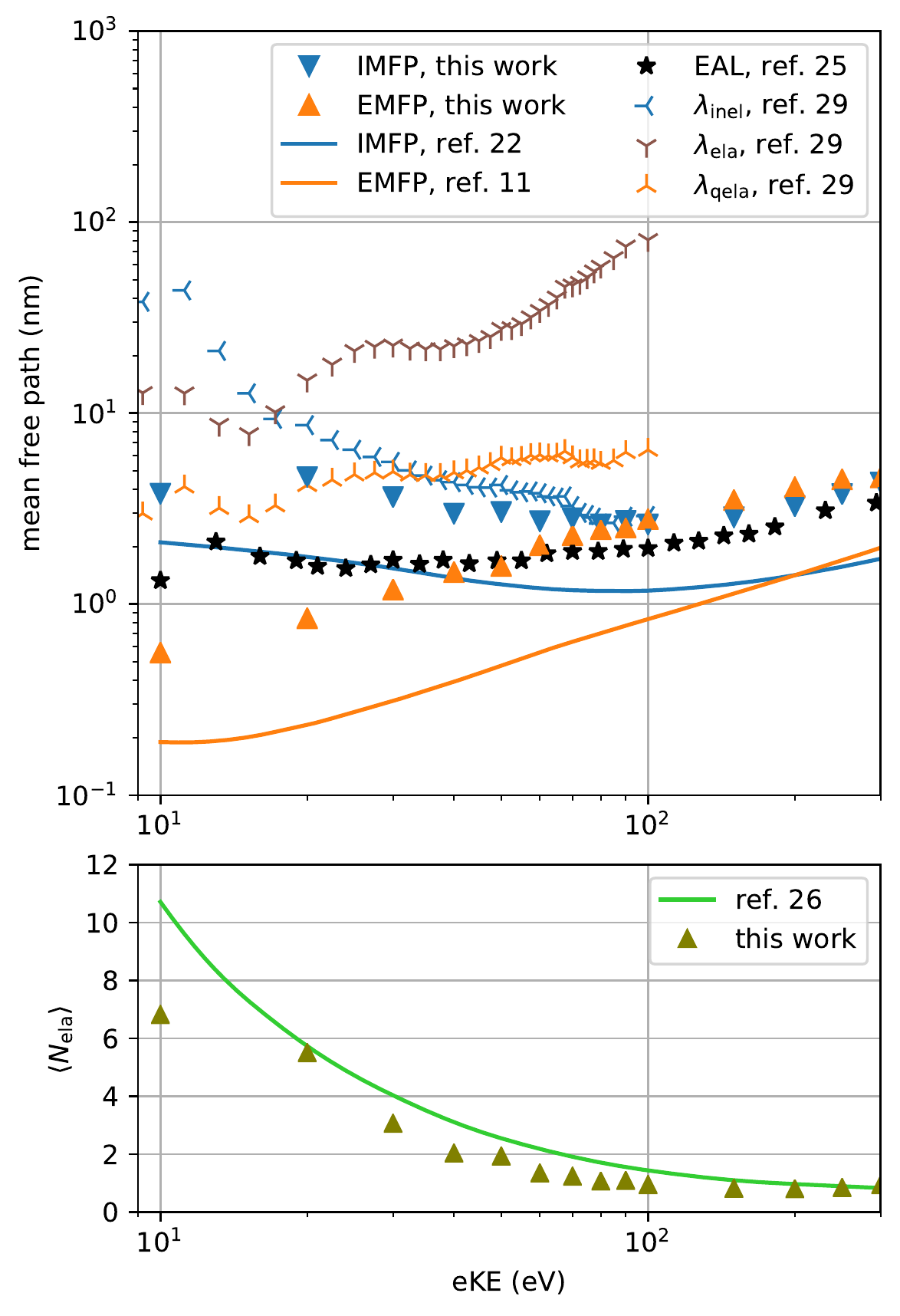}
    \caption{
    Top: Elastic (EMFP) and inelastic (IMFP) mean free paths from our simulation. 
    For comparison, the measured elastic mean free paths (EAL, extracted from 
    \citet{suzuki2014}) as well as the EMFP from \citet{tomita1997} 
    (corresponding to the integrated scattering cross sections for gas-phase 
    water from \citet{mark1995}), the IMFP from \citet{nguyen2018}, and 
    mean free paths obtained from \citet{michaud2003} for amorphous ice are shown.
    The elastic ($\lambda_{\rm ela}$), quasi-elastic ($\lambda_{\rm qela}$), and 
    inelastic ($\lambda_{\rm inel}$) mean free paths for \citet{michaud2003} are 
    obtained from the cross sections of the first column of Table 2, the sum of 
    the cross sections for all columns of Table 2 and of all but ``Others'' in 
    Table 3, and the cross sections ``Others'' in Table 3 of that reference, 
    respectively.
    Bottom: Average number of elastic collisions $\nela$. 
    The values from \citet{thurmer2013} are shown for comparison.
    }
    \label{fig:eimfp}
  \end{figure*}
  
We now turn to the discussion of the mean free paths obtained from the analysis of the experimental results.  
  The top panel of Figure \ref{fig:eimfp} shows the dependencies of the retrieved EMFP and 
  IMFP on the eKE.
  \blue{For comparison, the experimentally determined EAL used in the simulations is 
  shown, as well as literature values for the EMFP corresponding to the total 
  elastic scattering cross section for gas-phase water,\cite{mark1995,tomita1997} and
  the most recent theoretical estimation of the IMFP for liquid water\cite{nguyen2018}.}
  Figure \ref{fig:eimfp} also contains the mean free paths for purely elastic, 
  quasi-elastic (i.e. including vibrationally inelastic) and electronically 
  inelastic scattering determined by \citet{michaud2003} from experiments on 
  amorphous ice. 
  We note that the cross sections reported by \citet{luckhaus2017} are identical 
  to those of \citet{michaud2003} for all energies (from \unit[1.7]{eV} upwards) 
  reported in the latter.
  \blue{A comparison to the many other estimations of the IMFP can e.g.\ be found in 
  \citet{nguyen2018}.}
  
  The EMFP determined in our work (cyan triangles) is larger by a factor of 2-3 
  compared to previous estimates based on gas-phase data (cyan line). 
  It is smaller by a factor of  2-4 than the quasi-elastic MFP of 
  \citet{michaud2003} (cyan tripods). 
  We note that the purely elastic EMFP of \citet{michaud2003} (dark blue 
  tripods) is more than one order of magnitude longer than our EMFP and up to 
  2 orders of magnitude longer than the previous estimates.
  
  The IMFP determined in our work (magenta triangles) is found to be longer than
  the latest theoretical estimate \cite{nguyen2018} (magenta line) by a 
  nearly constant factor of $\sim2$. 
  Our IMFP is smaller than the electronic IMFP determined by 
  \citet{michaud2003} by up to one order of magnitude at \unit[10]{eV}, but 
  seems to merge with the latter around \unit[80]{eV}. 
  Our IMFP is larger than the experimental EAL by a factor of $\sim 3$ for an 
  eKE of \unit[$10$]{eV}, getting closer to the EAL for higher eKE, but always 
  remaining larger than the EAL, as required. 
  In our simulations, we find that the EAL becomes equal to the IMFP only if we 
  set the EMFP to be very long such that no elastic scattering happens at all.

  For eKE \unit[$\geq 60$]{eV}, both the IMFP and EMFP are larger than the EAL and comparable in magnitude. The EMFP eventually becomes somewhat larger than the IMFP for eKE \unit[$\geq 100$]{eV}, which is lower than the crossing point of the previous estimates around 200~eV. This does however not mean that elastic scattering becomes unimportant in the photoelectron experiments. A significant influence of elastic scattering at these high eKE can indeed be deduced from the experimental data because the PAD for the liquid phase remains different from that of the gas phase up to an eKE of \unit[$300$]{eV} \cite{thurmer2013}. We will return to this point below.

It is also interesting to note that our EMFP and IMFP values display a very similar energy dependence as the literature values, but are consistently higher by a factor of $\sim$2-3. This means that both sets of data, although completely independent, agree on the ratio of IMFP to EMFP, hence on $\nela$. Assuming that this ratio is correct, the vertical offset of our values of EMFP and IMFP can be traced back to the experimental EAL since, as we showed above, both the EMFP and the IMFP scale linearly with the input EAL for a fixed $\nela$. A reduction of the EAL by a factor of $\sim$3 would bring our EMFP and IMFP values into agreement with the previous estimates. What speaks in favor of such a correction are the expectations that (i) the dielectric formalism used to determine the IMFP from the measured energy-loss function of liquid water should be accurate above 100~eV \cite{nguyen2018} and (ii) the EMFP of liquid water might be expected to converge to that of isolated molecules for sufficiently high kinetic energies of a few hundred eV, where the electron scattering is dominated by core-shell electrons that are not affected by solvation and multiple scattering is negligible. What speaks against such a correction are the experimental results of \citet{michaud2003} on amorphous ice, which would then lie even much further away from those for liquid water.

  The bottom panel of Figure \ref{fig:eimfp} shows the average number of elastic
  collisions $\nela$ obtained from the present results (triangles) in comparison 
  to a previous analysis of the measured $\beta$-parameters for gas-phase and 
  for liquid water by \citet{thurmer2013}.
  Also in this case, the $\nela$ agree reasonably well. 
  This is, however, a coincidence. 
  In the analysis of the experimental results by \citet{thurmer2013} $\nela$
  was estimated by successive convolution of the PAD of the gas phase with a DCS originating from an educated guess until the measured PAD of the liquid was obtained.
  This analysis has two potential deficiencies:
  A minor problem is that a convolution with the DCS to account for multiple
  collisions is incorrect for the three-dimensional problem.
  Instead, for a DCS $D_{\rm S}(\theta)$ with polar angle $\theta$, the 
  angular distribution in the bulk $D^{(n+1)}(\theta)$ after $n+1$ collisions 
  is obtained from that after $n$ collisions as 
  \begin{align}
    D^{(n+1)}(\theta) = \int\limits_{0}^{\pi} \int\limits_{0}^{2\pi} D^{(n)}(\vartheta) D_{\rm S}(\vartheta') \sin(\vartheta) \, d\vartheta d \varphi \label{eq:con3d}
  \end{align}
  with 
  \begin{align}
    \cos \vartheta' = \cos(\vartheta) \cos(\theta) + \cos(\varphi) \sin(\vartheta) \sin(\theta).
  \end{align}
  The difference to a convolution is small but noticeable, as shown in the 
  supplemental material.
  
  \begin{figure*}[htbp]
    \includegraphics[width=0.8\textwidth]{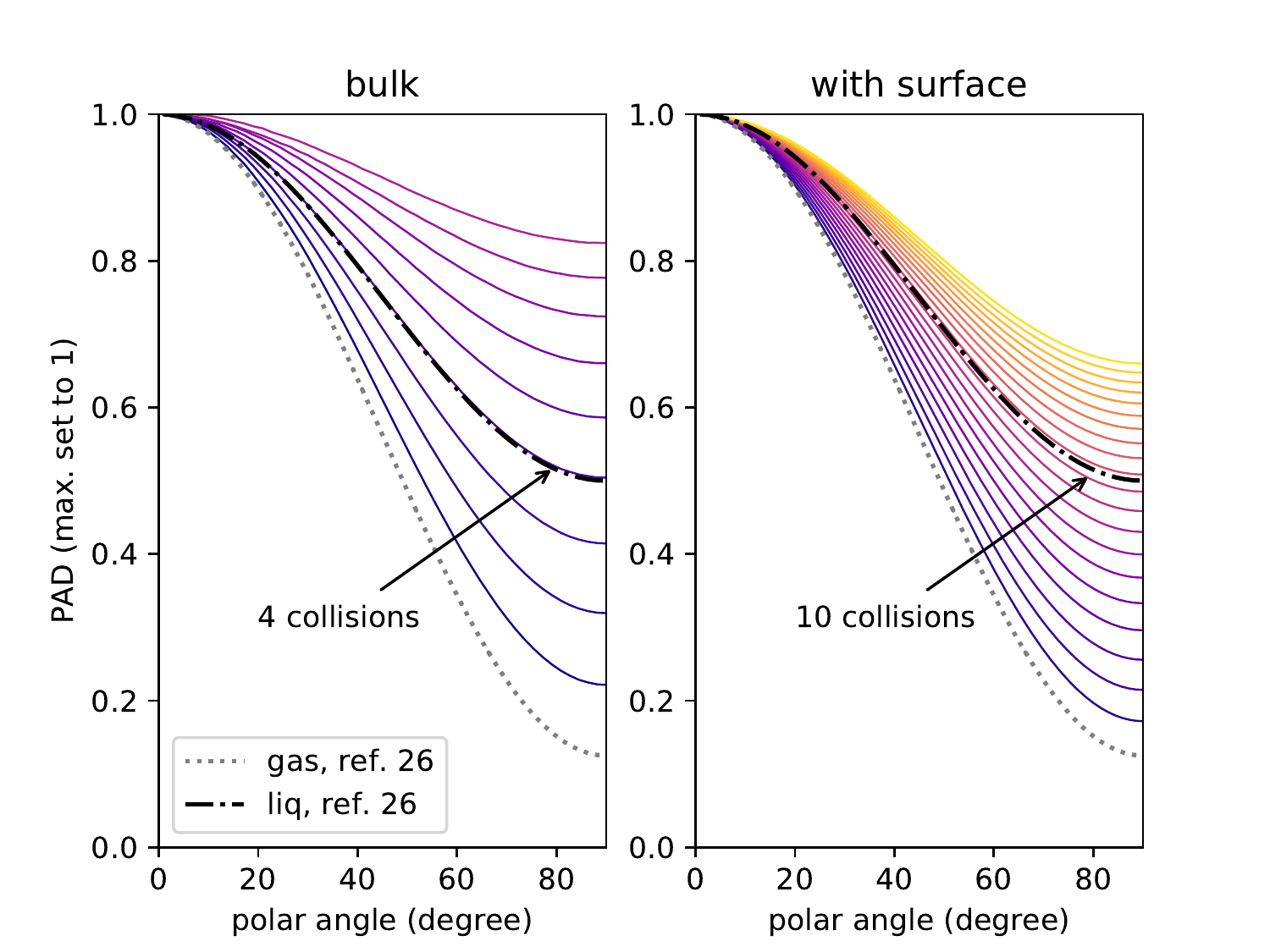}
    \caption{Photoelectron angular distribution (PAD, scaled such that the 
             maximum is 1) for liquid water after scattering in the bulk only
             (left) and determined outside the liquid (right).
             The kinetic energy of the scattered electron is \unit[20]{eV} and 
             the elastic mean free path is set to \unit[0.2]{nm}.
             Starting from the initial PAD (dotted line) each line above 
             corresponds to one additional collision. The dotted and 
             dash-dotted line correspond to measured PADs of the gas-phase 
             and the liquid-phase oxygen-1s-electrons from \citet{thurmer2013}.}
    \label{fig:comparison}
  \end{figure*}
  
  More important is the neglect of surface effects in the 
  analysis of \citet{thurmer2013}, as \eqref{eq:con3d} is only valid in the bulk.
  Due to of the relatively short EAL, in the experiment the contribution 
  from water molecules ionized close to the surface is important.
  If initially emitted in the direction of the detector, the corresponding electrons 
  cannot scatter as often as those ionized deeper inside the bulk.
  Hence, to generate the same final angular distribution electrons ionized 
  deeper in the liquid have to contribute, which requires more elastic 
  collisions and, consequently, a larger IMFP.
  
  The influence of the surface is illustrated in Fig.\ \ref{fig:comparison},
  where the PAD is shown after $n$ collisions as determined inside the 
  liquid and outside the liquid, for an eKE of \unit[20]{eV} and for an EMFP of
  \unit[0.2]{nm}.
  For the latter case, an average over the distance of the ionization site from 
  the surface was performed.
  While in the bulk about four collisions are needed to obtain the measured 
  distribution of the liquid, more than 10 collisions are needed if the surface 
  effect is taken into account.
  The inclusion of the surface in the simulation is thus crucial for the 
  analysis of the measurement.
  Nevertheless, the difference of the $\nela$ shown in 
  Figure \ref{fig:eimfp} is small, because we find larger EMFP values compared 
  to the values used by \citet{thurmer2013}.
  Before concluding, we note that the EMFP and IMFP values determined in our 
  work are related to physical \emph{ab-initio} DCS used in our simulations. 
  This must be kept in mind, both when comparing our results with other sources, 
  some of which assume isotropic scattering and the associated transport cross 
  sections\cite{jablonski1999}, and when using our results in simulations.
  
  In conclusion, we have introduced a novel, straightforward, and general
  method to obtain the EMFP and IMFP of electrons scattering in liquids
  that is based on the simulation of two types of experiments.
  Our approach is possible due to the discovery of a rapid convergence of the DCS
  of water clusters with cluster size, which provides the first 
  \textit{ab-initio} molecular-level description of electron scattering in liquid water.
  \red{It is found that to be consistent with the measured EAL and PAD of soft-X-ray photoelectron 
  experiments, the liquid-phase IMFP must be larger than the latest estimate from the dielectric formalism \cite{nguyen2018} and the liquid-phase EMFP must be larger than that obtained from scattering data of gas-phase water \cite{tomita1997}, respectively.}
  We also find that on average only few elastic collisions are possible 
  before inelastic scattering occurs even for relatively low eKE, but that 
  elastic scattering remains important for eKE values up to \unit[300]{eV}. In the studied photoelectron experiments, the importance of elastic scattering is accentuated by the surface sensitivity resulting from the EAL. 
  
  The reliability of the mean free paths obtained in our work depends on the accuracy of the experimental input as well as on
  the accuracy of the DCS used for bulk liquid water.
  The sensitivity of the retrieved mean free paths on the input parameters is illustrated in Figure S3 of the supplemental material.
  Hence, further experimental studies would be highly desirable and we hope that 
  our analysis will stimulate additional measurements, also on other liquids.
  The program code used for this work is available as Python module at
  \url{https://gitlab.com/axelschild/CLstunfti} and can be used to obtain 
  improved values for the EMFP and IMFP when additional experimental or theoretical data becomes available.
  We expect that our method to determine the mean free paths will become a 
  powerful tool for understanding electron scattering in liquid media, when 
  combined with accurate experimental and theoretical input data. 
  This opens the door for gaining better knowledge of the properties of 
  liquid water, for the exploration of the scattering properties of other 
  liquids, and for the interpretation of attosecond time-resolved measurements 
  in the liquid phase \cite{jordan15a,jordan18b}.
  
  \acknowledgement{
    AS thanks Jakub Koc\'ak (ETH Z\"urich) for helpful discussions, Denis Jelovina 
    (ETH Z\"urich) for providing some structures of the considered water clusters, 
    and the Swiss National Science Foundation for supporting this research with an 
    Ambizione grant.
  }

  \bibliography{bib}{}
  
\end{document}